\documentstyle[twocolumn,aps,prb,epsf]{revtex}
\begin{document}
\newcommand{\beq}{\begin{equation}}
\newcommand{\eeq}{\end{equation}}

\title{Heat capacity of mesoscopically disordered superconductors:
implications to MgB$_{2}$}
\author{A. M. Gabovich$^1$, A. I. Voitenko$^1$, Mai Suan Li$^2$,
H. Szymczak$^2$, and M. Pekala$^3$}

\address{$^1$ Crystal Physics Department, Institute of Physics, NUAS,
Prospect Nauki 46, 03022 Kiev-22, Ukraine  \\
$^2$Institute of Physics, Polish Academy of Sciences,
Al. Lotnikow 32/46, 02-668 Warsaw, Poland\\
$^3$Department of Chemistry, University of Warsaw,
Al. Zwirki i Wigury 101, PL-02-089 Warsaw, Poland}



\address{
\centering{
\medskip\em
{}~\\
\begin{minipage}{14cm}
Electronic heat capacity $C(T)$ was calculated for a mesoscopically
disordered $s$-wave superconductor treated as a spatial ensemble of
domains with a continuously varying superconducting properties. The
domains are assumed to have sizes $L>\xi _{0}$, where $\xi _{0}$ is the
coherence
length. Each domain is characterized by a certain critical temperature $%
T_{c0}$ in the range $[0,T_{c}]$. The averaging over the superconducting
gap distribution leads to $ C(T) \sim
T^{2}$ for low $T$, whereas the specific heat anomaly at $T_{c}$ is
substantially smeared. The results explain well the $C(T)$ data for
MgB$_{2}$, where a multiple-gap structure is observed.
{}~\\
{}~
\end{minipage}
}}


\maketitle


An unexpected discovery of relatively high-temperature, high\textrm{-}$T$,
superconductor MgB$_{2}$ with a critical temperature $T_{c}\approx 40\,%
\mathrm{K}$ \cite{nagamatsu01:63} have cast a considerable doubt on the
validity of the opinion that high $T_{c}$'s are appropriate to substances
with a spin-fluctuation-driven Cooper pairing and, consequently, with a
predominantly $d_{x^{2}-y^{2}}$-wave symmetry of the superconducting order
parameter. Indeed, an obvious absence of magnetic ions, a considerable
isotopic effect \cite{budko01:1877}, and Bardeen-Cooper-Schrieffer-like
(BCS-like) coherent peaks in optical conductivity \cite{pronin01:097003}\
and spin-lattice relaxation \cite{kotegawa01:127001} are indicative of the
conventional \textit{s}-wave pairing in MgB$_{2}$. As for the
electron-phonon background of superconductivity, it also seems highly
probable, although a fairly exotic multiple-gap scenario is needed to
reconcile the available data (see, e.g., Ref. \onlinecite{liu01:087005}).
It is very remarkable that the multiple-gap conventional Cooper pairing is
directly found in a number of point-contact, tunneling and Raman
measurements \cite{buzea:0108265}. It is even more important that the
distribution of gaps may be rather broad \cite{laube01:296}\ and
spatially-resolved \cite{sharoni01:503}, although much controversy exists
over the number and widths of gaps in the electron density of states
(DOS).

In actual truth, the order parameter symmetry for MgB$_{2}$ is not
unambiguously determined. The low-$T$, asymptotics of the magnetic field
penetration depth $\lambda (T)$ was shown by muon spin-rotation \cite
{panagopoulos01:094514} and optical \cite{pronin01:097003}\ measurements
to be a power-law one. This was interpreted as either an unconventional
superconductivity or at least as a highly anisotropic \textit{s}-wave
pairing. Thermodynamic measurements might be decisive in determining the low-%
$T$ symmetry-based superconducting properties of MgB$_{2}$ because the
minority phases or grain boundaries do not affect the results
substantially
in contrast to, e.g., transport phenomena. The electronic heat capacity, $%
C(T)$, behavior near $T_{c}$ is also of great importance to elucidate the
nature of superconductivity here. And, indeed, there were a lot of
specific heat investigations for MgB$_{2}$ performed by various groups
\cite {junod01:179}.

The main features of the data for $C(T)$ are (i) small values of the phase
transition anomaly $\Delta C=C_{s}-C_{n}$ at $T_{c}$ \cite
{junod01:179,wang01:179,kremer:0102432,yang01:167003}\ in comparison to
the BCS case \cite{abrikosov87:book-en}, when the ratio $\mu =\Delta
C/[\gamma _{S}(T_{c})T_{c}]$ is equal to $\mu _{\mathrm{BCS}}=12/[7\zeta
(3)]$; and (ii) deviations from the asymptotic BCS behavior at $T\ll
T_{c}$
\begin{equation}
C_{s}^{\mathrm{asympt}}(T)=N(0)\left( \frac{2\pi \Delta _{0}^{5}}{T^{3}}%
\right) ^{1/2}\exp \left( -\frac{\Delta _{0}}{T}\right) .
\label{CsBCSasy}
\end{equation}
Here $\gamma _{S}$ is a Sommerfeld constant, the subscripts $s$ and $n$
correspond to the superconducting and normal states, respectively, $N(0)$
is the electron DOS at the Fermi level, $\Delta _{0}$ is the energy gap
value at $T=0$, $k_{B}=\hbar =1$. The deviations from Eq. (\ref{CsBCSasy})
may be twofold: power-law-like $\sim T^{2}$ \cite{wang01:179} and of
the form $\sim \exp ( -\frac{A}{T}) \,$\cite
{yang01:167003,bouquet01:047001}, where the constant $A$ is much less than $%
\frac{\pi }{\gamma }T_{c}\approx 1.76\,T_{c}$, as it should be in the
weak-coupling superconductor \cite{abrikosov87:book-en}, $\gamma
=1.78\ldots $ is the Euler constant. Thus, the raw specific heat data do
not give definite answers to the problems of the order parameter symmetry
and the underlying mechanisms of superconductivity.

In this article on the basis of the experimentally proved \textit{%
distribution} of energy gaps we show that \textit{both} main features of $%
C_{s}(T)$ can be explained by the conventional \textit{s}-wave
superconductivity, so that these data can be easily reconciled with other
observations \cite{pronin01:097003,kotegawa01:127001}. The adopted
approach, being the outgrowth of the earlier one \cite{gabovich99:7465},
is phenomenological because the origin of the gap distribution is not
known precisely. However, in accordance with tunneling data
\cite{sharoni01:503}, the gap distribution is considered to occur
\textit{in the real space} rather than in the $\mathbf{k}$-space, as was
suggested, e.g., in Refs. \onlinecite {junod01:179,wang01:179}. The
theoretical description of such spatially disordered superconductors
depends on the ratio between the characteristic superconducting domain
size $L$ and the coherence length $\xi _{0}$ \cite {larkin71:2147}. If
$L>\xi _{0}$, superconducting properties are determined by local values of
the order parameter $\Delta $. In the case of MgB$_{2}$ there is a large
scatter of $\xi _{0}$, inferred from different experiments and for
\textit{different kinds} of samples \cite{buzea:0108265}, so that we may
estimate this quantity as lying in the range from $25\,$\textrm{\AA } to
$65\,$\textrm{\AA }. This dispersion of $\xi _{0}$ qualitatively
correlates with the broad spectra of gaps in tunnel and point-contact
spectra \cite {sharoni01:503,laube01:296,junod01:179,buzea:0108265}.



Let us examine a $T$-independent configuration of mesoscopic domains, with
each domain having the following properties:

(A) at $T=0$ it is described by a certain superconducting order parameter $%
\Delta _{0}\leq \Delta _{0}^{\max }$;

(B) up to a relevant critical temperature $T_{c0}(\Delta
_{0})=\frac{\gamma }{\pi }\Delta _{0}$, it behaves like an isotropic BCS
superconductor, i.e. the superconducting order parameter $\Delta (T)$ is
the M\"{u}hlschlegel function $\Delta (T)=\Delta _{\mathrm{BCS}}(\Delta
_{0},T)$ \cite {abrikosov87:book-en}; and the electronic specific heat is
characterized in this interval by the function $C_{s}(\Delta ,T)$;

(C) at $T>T_{c0}$ it transforms into the normal state, and the relevant
property is $C_{n}(T)$.

At the same time, the values of $\Delta _{0}$ scatter for various domains.
The current carriers move freely across domains and inside each domain
acquire appropriate properties. The adopted picture is especially suitable
for superconductors with small coherence lengths $\xi _{0}$ \cite
{junod01:179}.

In other words, each domain above its $T_{c0}$ is in the normal phase and
its specific heat is \cite{abrikosov87:book-en}
\begin{equation}
C_{n}(T)=\frac{\pi ^{2}}{3}N(0)T.  \label{Cn}
\end{equation}
For simplicity we restrict ourselves to the situation when the whole
sample above $T_{c}$ is electronically homogeneous, i.e. is characterized
by a common approximately constant $N(0)$ value. Below $T_{c0}$ for a
given mesoscopic domain, a corresponding isotropic gap appears on the
Fermi surface. The microscopic background of the assumed scatter in
$T_{c0}$'s may be diverse but ultimately manifests itself as a variation
either of the electron-phonon interaction magnitude or of local values of
the Coulomb pseudopotential.

In the framework of our phenomenological approach, superconductivity (if
any) inside a chosen domain is described by the relevant parameters
$\Delta _{0}$ and $T_{c0}$. They are bounded from above by $\Delta
_{0}^{\max }$ and $T_{c}$, respectively. These $\Delta _{0}$'s may or may
not group around a certain crowding value $\Delta _{0}^{\ast }$ depending
on the sample texture. The existence of such two possibilities is in
accordance with the varied data for MgB$_{2}$ \cite
{sharoni01:503,laube01:296,junod01:179,buzea:0108265,bouquet:0107196}. The
specific gap distribution is described by the function $f_{0}(\Delta
_{0})$.

Thus, for all $T$ in the interval $[0,T_{c}]$, where $T_{c}=\max
\,T_{c0}$, the superconducting sample consists of superconducting (s) and
nonsuperconducting (n) grains more or less homogeneously distributed over
the sample volume.

The measured $C_{s}(T)$ is an averaged sum of contributions from both
phases
\begin{equation}
\left\langle C(T)\right\rangle =\left\langle C_{n}(T)\right\rangle
+\left\langle C_{s}(T)\right\rangle ,  \label{<C>}
\end{equation}
which depends on the distribution function $f(\Delta ,T)$ of
superconducting domains, and on the fraction $\rho _{n}(T)$ of the normal
phase \cite {gabovich99:7465}
\begin{equation}
\left\langle C_{n}(T)\right\rangle =C_{n}(T)\rho _{n}(T),  \label{<Cn>}
\end{equation}
\begin{equation}
\left\langle C_{s}(T)\right\rangle =\int_{0}^{\Delta ^{\mathrm{\max }%
}(T)}C_{s}(\Delta ,T)\,f(\Delta ,T)\,d\Delta .  \label{<Cs>}
\end{equation}
Here $\Delta ^{\mathrm{\max }}(T)=\Delta _{\mathrm{BCS}}(\Delta _{0}^{\max
},T)$ and $f(\Delta ,T)$ is a result of the thermal evolution of the
initial (at $T=0$) distribution function $f_{0}(\Delta _{0})$. It is
convenient to normalize all temperatures by $T_{c}$ and all energy
parameters by $\Delta _{0}^{\max }$: $t=T/T_{c},\delta =\Delta /\Delta
_{0}^{\max }$ with relevant indices retained, and to consider $C_{s}(T)$
and $C_{n}(T)$ together with
their averaged counterparts, normalized by the $C_{n}(T_{c})$ value, i.e., $%
c_{s,n}(t)=C_{s,n}(T)/C_{n}(T_{c})$. Then one can easily find that for
each domain, characterized by the parameter $\delta _{0}$ at $t=0$, the
dimensionless heat capacity is either
\begin{equation}
c_{n}(t)=t,\,t>\delta _{0}  \label{cn}
\end{equation}
or
\begin{equation}
c_{s}(t)=\delta _{0}c_{\mathrm{BCS}}\left( \frac{t}{\delta _{0}}\right)
,\,t<\delta _{0},  \label{cs}
\end{equation}
where $c_{\mathrm{BCS}}(x)$ is a well-known normalized heat-capacity
function for a standard BCS superconductor \cite{abrikosov87:book-en}. For
a surmised domain ensemble a distribution function $f(\Delta ,T)$ for
finite $T $ is defined by the formula
\begin{equation}
f(\Delta ,T)\,d\Delta =f_{0}(\Delta _{0})\,d\Delta _{0}.
\end{equation}
Then the dimensionless heat capacity takes the form
\begin{equation}
\left\langle c_{s}(t)\right\rangle =\int_{t}^{1}\,c_{\mathrm{BCS}}\left(
\frac{t}{\delta _{0}}\right) \,f_{0}(\delta _{0})\,\delta _{0}\,d\delta
_{0}.  \label{<cs>}
\end{equation}
Introducing a new variable $z=t/\delta _{0}$ and expanding the function $%
f_{0}\left( t/z\right) $ into a series we arrive at the proper low-$t$
asymptotics
\begin{equation}
\left\langle c_{s}(t\rightarrow 0)\right\rangle
\; = \; t^{2}\,\int_{0}^{1}\,\frac{%
dz}{z^{3}}f_{0}\left( 0\right) \,c_{BCS}(z)\approx
2.45\,f_{0}\left( 0\right) \,t^{2}.  \label{csasy}
\end{equation}

The $t$-dependence of the next term in the expansion for $\left\langle
c_{s}(t)\right\rangle $ can be estimated in the limit $t\rightarrow 0$ by
substitution of the normalized expression (\ref{CsBCSasy}) for $c_{\mathrm{%
BCS}}(z)$. \ It turns out that this expression decreases as $O[t^{5/2}\exp (-%
\frac{\pi }{\gamma t})]$.

Now, in the same low-$T$ region let us take a look at the contribution $%
\left\langle c_{n}(t)\right\rangle $ of the continuously expanding normal
phase. At any $T$, all domains with $\Delta _{0}<\frac{\pi }{\gamma }T$
(i.e. $\delta _{0}<t)$ are nonsuperconducting, with the total normal phase
fraction being
\begin{equation}
\rho _{n}(t)=\rho _{n}(0)+\int_{0}^{t}f_{0}(\delta _{0})\,d\delta _{0}.
\label{ron(T)}
\end{equation}
For simplicity, below we restrict ourselves to the case when all domains at $%
t=0$ are superconducting, i.e. $\rho _{n}(0)=0$. A generalization to the
case $\rho _{n}(0)\neq 0$ is obvious: at each temperature there exists an
additional contribution from the normal phase. Then the function $%
f_{0}(\delta _{0})$ should be normalized by $1-\rho _{n}(0)$, and all
averaging-driven effects would accordingly decrease. Moreover, if $\rho
_{n}(0)\neq 0$, the observed heat capacity $\langle c(t)\rangle $ must
include an extra linear contribution $\rho _{n}(0)t$ in the true
superconducting state exhibiting the Meissner effect.

As for the second term in Eq. (\ref{ron(T)}), the approximation of $%
f_{0}(\delta _{0})$ by its limiting value $f_{0}(0)$ demonstrates that the
main temperature-dependent contribution to $\rho _{n}(t)$ is \textit{linear }%
in $t$. Since $c_{n}(t)$ is also a linear function of $t$, the apparent
contribution $\left\langle c_{n}(t)\right\rangle $ of the normal phase to
the resulting specific heat $\langle c(t)\rangle $ is quadratic in $t$ for
small $t$, similarly to $\left\langle c_{s}(t)\right\rangle $. Thus, in
the suggested model of the disordered superconductor with a broad\
continuous spatial distribution of domains, possessing different
$T_{c}$'s, normal and
superconducting contributions to thermodynamical quantities are \textit{%
functionally indistinguishable} from each other.


In addition to the low-$T$ asymptotics the overall $T$-dependence of the
heat capacity $C$ up to $T_{c}$ is also of considerable interest.
Especially
important is to trace the smearing of the anomaly $\Delta C$ by the \textit{%
same }effect of disorder which leads to the transformation of the
intrinsic exponential low-$T$ behavior of $C_{s}(T)$ into a power-law one.
These objectives were met by numerical calculations.

For this purpose, a Gaussian model distribution function $f_{0}^{G}(\delta
_{0})$ was used:
\begin{equation}
f_{0}^{G}(\delta _{0}^{{}})\propto \exp \left[ -\frac{\left( \delta
_{0}^{{}}-\delta _{0}^{\ast }\right) ^{2}}{2d^{2}}\right] .
\label{f-gauss}
\end{equation}
The parameter $\delta _{0}^{\ast }$ designates the peak position, which
may vary from 0 to 1. By changing the parameter $d$ we control the
dispersion of the domain superconducting properties. Nevertheless, for any
$d$ $\ $the function $f_{0}^{G}(\delta _{0})$ does not vanish in the limit
$\delta _{0}=0 $ and its Taylor series begins with a constant as the main
term. Only
for highly improbable distribution functions, when simultaneously $%
f_{0}(\delta _{0})$ extends to $\delta _{0}=0$ and matches the condition $%
f_{0}(\delta _{0}=0)=0$, the Taylor series may begin with the next term
resulting in the asymptotics $C_{s}(T) \sim T^{3}$.

\begin{figure}
\epsfxsize=3.2in
\vspace{-0.2in}
\centerline{\epsffile{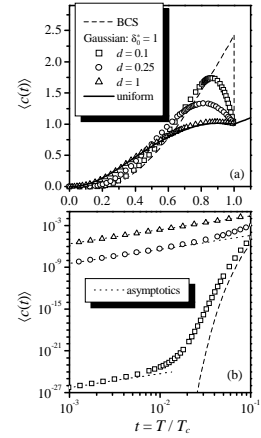}}
\vspace{-1.5in}
\caption{(a) Temperature dependences of normalized total electronic heat
capacity $\left\langle c(t)\right\rangle $ in comparison with the
BCS-dependenceof superconducting phase fraction. Gaussian distributions
with $\protect\delta _{0}^{\ast }=1$. (b) Low-temperature portions of the
relevant curves on log-log scale together with their $t^{2}$-asymptotics.
} \label{fig:Cat1}
\end{figure}

In Fig. \ref{fig:Cat1}
the dependences $\langle c(t)\rangle $ are depicted in the panel (a) for $%
\delta _{0}^{\ast }=1$ and different dispersion values $d$. A substantial
spreading of the anomaly $\Delta C$ readily seen in Fig. \ref{fig:Cat1}
seems quite natural in view of the results for MgB$_{2}$ \cite
{junod01:179,wang01:179,kremer:0102432,yang01:167003}. However, the
concomitant superposition of various domain contributions distorts the
whole curves $C_{s}(T)$ and $C(T)$, which is much less trivial. This very
superposition leads for low $T$ to the power-law behavior, the asymptotics
of which was analyzed above. The low-$T$ parts of the curves $\langle
c(t)\rangle $ are displayed on the log-log scale in the panel (b). Dotted
straight lines correspond to the pertinent $T^{2}$-asymptotics for each
curve. It is clear that the validity range of the asymptotics extends with
the increase of $d$. Although intervals where the $T^{2}$-approximation
holds good exist for any $d$, for small $d$ it is merely of academic
interest, because both temperatures and heat capacities become too tiny to
be experimentally significant. On the other hand, for higher $T$ in this
case the averaged dependences $\langle c(t)\rangle $ lie rather close to
the exponential curve inherent to the BCS theory (the dashed curve). Such
transitional parts of the dependences $\langle c(t)\rangle $ describe well
the exponential low-$T$ behavior for some samples of MgB$_{2}$ \cite
{yang01:167003,bouquet01:047001} with smaller exponents than in the BCS
case.

For large $d$, when the Gaussian distribution function $f_{0}^{G}(\delta
_{0})$ becomes almost uniform (such a random dense, although
quasi-discrete, distribution of gaps was found in point-contact spectra
\cite{laube01:296}), the quadratic asymptotics are valid at least up to
$t=0.1$ (for the uniform distribution $f_{0}^{U}(\delta
_{0})=\mathrm{const}$ the relative error of the $t^{2}$-asymptotics is
$\approx 0.6\%$ at $t=0.1$ and $\approx 5\%$ at $t=0.2$), which
agrees with the measurements \cite{wang01:179}. For intermediate $d$ the
experimental data in the relevant $T$-range may be
satisfactorily represented by power-law curves $C(T) \sim T^{n}$ with $%
n \ge 2$.

One can make another important conclusion from the numerical data shown in
Fig. \ref{fig:Cat1}. A one-parameter fitting explains \textit{both} the
smearing of the heat-capacity anomaly at $T_{c}$ and the appearance of the
power-law asymptotics. The latter reproduces the results appropriate to
superconductors with order parameters of $d_{x^{2}-y^{2}}$-wave \cite
{won01:375} or extended \textit{s}-wave with uniaxial anisotropy \cite
{junod01:179,wang01:179}\ symmetry. The patterns displayed in these
figures
explain well the experimental heat capacity dependences $C(T)$ for MgB$_{2}$%
, which demonstrate power-law behavior for lowest attainable $T$ \cite
{junod01:179,wang01:179} or above the exponential low-$T$ tail \cite
{bouquet01:047001}. At the same time, the reduction of the anomaly $\Delta
C$ at $T_{c}$ with the increase of $d$, traced in Fig. \ref{fig:Cat1}(a),
adequately describes the $\Delta C$ magnitudes inferred from the analysis
of the observed total heat capacity of MgB$_{2}$, making allowance for
crystal lattice and impurity components. Namely, $\mu \approx 1.13$
\cite {yang01:167003}, $0.82$ \cite{junod01:179,wang01:179}, $0.7$ \cite
{kremer:0102432}, so that the experimental specific heat jump is
substantially smaller than the BCS value $\mu _{BCS}$.

To summarize, we presented a phenomenological model of the disordered
$s$-wave superconductor with a random domain network possessing
continuously varying superconducting properties. The characteristic domain
size $L$ is supposed to exceed the superconducting coherence length $\xi
_{0} $. The spatially averaged electronic heat capacity $\langle
C(T)\rangle $ is
calculated. It is shown that its low-$T$ asymptotics is a power-law one $%
\sim T^{2}$, whereas the anomaly $\Delta C$ at $T_{c}$ is
simultaneously smeared. These are just the features appropriate to the
heat capacity of MgB$_{2}$. Although we do not know the exact nature of
the partition into domains in this compound, its very existence
undoubtedly manifests itself in Raman, point-contact and tunnel spectra
\cite {buzea:0108265,laube01:296,sharoni01:503}. One may speculate that
the multiple-gap superconductivity originates from some kind of a phase
separation rather than from the existence of several groups of current
carriers in the same volume \cite{liu01:087005}. The presented theory may
be
also invoked to explain low-$T$ properties of cuprates \cite{gabovich99:7465}%
, although the microscopical background of the multi-gapness may be quite
different in both cases.

\begin{acknowledgments}

A. M. G. is grateful to the Mianowski Foundation for support of his visit
to Warsaw University. The work was supported by the Polish agency KBN
(Grants No 2P03B-146-18 and 7T08A-028-20).

\end{acknowledgments}



\end{document}